\documentclass[11pt]{article}
\usepackage{graphicx} 
\usepackage{caption}
\usepackage[margin=1in]{geometry}
\usepackage{authblk}
\usepackage[hyperfootnotes=false,hidelinks]{hyperref}
\usepackage{booktabs}
\usepackage{array}
\usepackage{longtable}
\usepackage{amsmath,amssymb}
\usepackage[english]{babel}
\usepackage[T1]{fontenc}

\newcommand{\equalcontrib}{\textsuperscript{\ensuremath{\dagger}}}
\newcommand{\workdone}{\textsuperscript{\ensuremath{\ddagger}}}

\title{\textbf{Evaluating SAT Solver Metrics as Predictors of Human-Perceived Nonogram Difficulty}}
\author[1]{\textbf{Changdao He}\equalcontrib\workdone}
\author[2]{\textbf{Yibing Ju}\equalcontrib}
\author[2]{\textbf{Jonathan Calver}}
\author[2]{\textbf{Alice Gao}}

\affil[1]{Department of Computing Science, University of Alberta}
\affil[2]{Department of Computer Science, University of Toronto}
\affil[ ]{\texttt{mhe14@ualberta.ca} \quad \texttt{bing.ju@mail.utoronto.ca} \quad \texttt{calver@cs.toronto.edu} \quad \texttt{ax.gao@utoronto.ca}}

\begin{document}

\maketitle

\begingroup
\renewcommand{\thefootnote}{\fnsymbol{footnote}}
\footnotetext[2]{These authors contributed equally.}
\footnotetext[3]{Work done while at the University of Toronto.}
\endgroup

\begin{abstract}

Algorithmic solver effort is often assumed to align with perceived puzzle difficulty, but this assumption is rarely tested against human solving data. We evaluate this assumption for Nonograms, a popular logic puzzle similar to Sudoku in which numeric clues along each row and column determine a unique solution grid. We formulate Nonograms as a constraint satisfaction problem and solve them using existing SAT solvers. We then conduct a user study in which we collect data on both participant interactions and reported difficulty. We find that neither participants' reported difficulty nor their behavioural signals correlate meaningfully with SAT solver metrics; however, we find evidence that expertise moderates the relationship between solver metrics and reported difficulty. In this process, we uncover distinct, recurring solving strategies that indicate human preference for complex propagation, diverging from solver-measured complexity.

\end{abstract}

\newpage

\section{Introduction}

\begin{figure}[ht]
    \centering
    \includegraphics[width=0.45\linewidth]{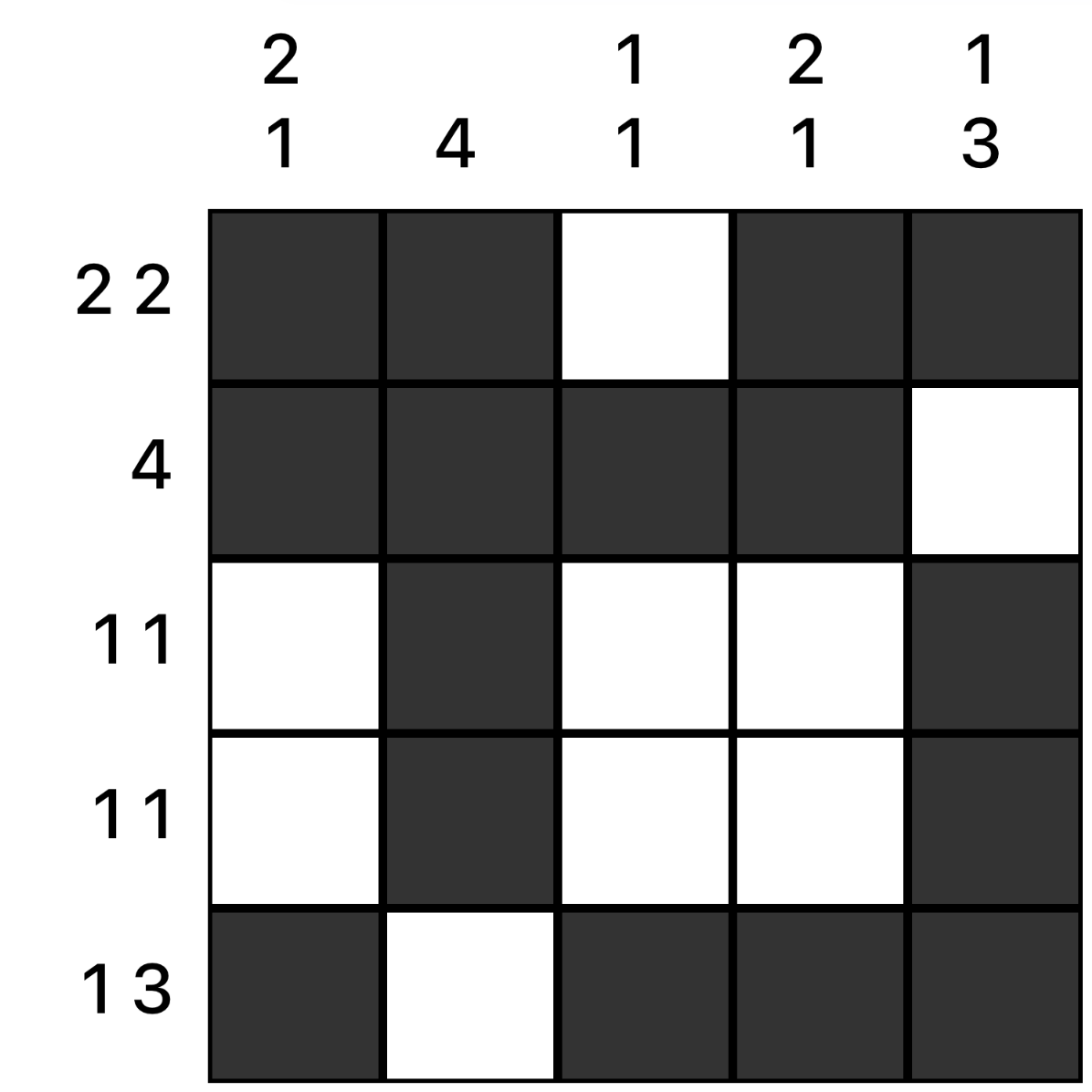}
    \caption{Example Nonogram. Numbers beside each row and column indicate the lengths of consecutive filled-cell runs, in order. Filled cells form the unique solution.}
    \label{fig:nonogram-example}
\end{figure}

Logic puzzles have been studied extensively; in particular, the efficacy of solver metrics at predicting difficulty has often been investigated for grid-based games such as Sudoku (for an overview, see \cite{pelanek_difficulty_2014}). Here, we study Nonograms, which are logic puzzles in which the clues beside each row and column specify the lengths of consecutive runs of filled cells. For example, the clue ``$1\ 3$'' denotes a run of one filled cell followed by at least one empty cell which is in turn followed by a run of three filled cells. The player must determine which cells are filled so that all row and column clues are satisfied simultaneously. Figure~\ref{fig:nonogram-example} shows a small example and its unique solution.

Multiple prior studies have examined how algorithmic solvers can be used to measure puzzle difficulty \cite{batenburg_constructing_2009, roucairol_generating_2024}, but most lack validation on human data. Motivated by this, we ask two research questions.

\begin{itemize}
    \item \textbf{RQ1}: To what extent do SAT solver statistics correlate with participant-reported puzzle difficulty?
    \item \textbf{RQ2}: To what extent do SAT solver statistics correlate with human solving behaviour?
\end{itemize}

We formulate Nonogram solving as a Boolean satisfiability (SAT) problem and solve the resulting instances using conflict-driven clause learning (CDCL) solvers, from which we derive solver statistics including decisions, propagations, and conflicts. We then conduct a behavioural study that records participant interactions and self-reported difficulty. We expected greater solver effort to correlate with higher reported difficulty. Further, we expected specific solver metrics to align with certain behaviour; for example, the number of conflicts encountered by the solver may correlate with the number of errors made by a human participant, and the number of decisions encountered by the solver may correlate with the number of pauses made by a human participant.

In this exploratory study, we find little evidence that SAT solver statistics positively associate with participant-reported difficulty or with behavioural signals--an initially discouraging result for the specific solver metrics, encoding, and puzzle sample examined. Participant trajectories and survey responses further suggest that starting footholds and constraint-propagation chains may be relevant to human judgments but are not directly represented by the SAT statistics considered.

\section{Related Work}

\subsection{Algorithms and Solvers}

Many algorithms have been proposed for solving grid-based puzzle games. For Nonograms in particular, the methods employed include genetic algorithms \cite{tsai_solving_2011}, depth-first search and permutations generation \cite{wieckowski_algorithms_2021}, and a heuristic algorithm combined with a neural network \cite{buades_rubio_solving_2024}.

Constraint-based formulations of Nonograms include a 2-SAT-based reasoning framework \cite{batenburg_reasoning_2008}, an integer linear programming formulation for colored Nonograms that generalizes an earlier black-and-white model \cite{mingote_colored_2009}, and a CSP-based solver that decomposes the puzzle into separate row and column CSPs before combining them into a global CSP \cite{aramian_solving_2025}. Human-oriented solver approaches have also been proposed, including line-by-line reasoning and deterministic guessing \cite{batenburg_constructing_2009, roucairol_generating_2024}. Unlike these decomposition-based CSP formulations, our formulation encodes the complete puzzle as a single CNF instance with explicit block-start variables, which we detail in Section~\ref{sec:formulation}.

\subsection{Puzzle Difficulty}

Puzzle difficulty, and Nonogram difficulty in particular, has received considerable attention. In terms of mathematical difficulty, Nonograms are shown to be NP-hard \cite{ueda_np-completeness_1996}, and the difficulty of the inference problem (co-NP-complete) is largely determined by the density of filled cells \cite{foote_nonogram_2025}. However, we are interested in the human-perceived difficulty of Nonogram puzzles, often studied in the context of the automatic generation of puzzles. \cite{de_kegel_procedural_2020} provide an overview of procedural generation of puzzles in a broader context, emphasizing that difficulty is not a single intrinsic property, but instead reflects the interaction between problem structure, solving processes, and evaluation criteria. For specific grid-based logic puzzles, it is common to identify features likely to correlate with difficulty and derive weighted difficulty measures, using user ratings \cite{kartal_data_2016} or a proxy such as solve time \cite{wang_rating_2012}. Computational measures such as informational entropy \cite{chen_entropy_2023} have also been studied in relation to user feedback. On the other hand, difficulty measures have been composed from metrics from human-like solvers \cite{batenburg_constructing_2009, roucairol_generating_2024} but have not been validated against actual human subjects.

We note that solve time and perceived difficulty have been used interchangeably in much of the existing literature \cite{wang_rating_2012, naji_sudoku_2024, wang_dataset_2024}. However, more recent work suggests that while the two measures track each other reasonably well in general, they decouple around insight moments \cite{moroshkina_aha_2024}. Therefore, we aim to obtain a more holistic understanding of human perception of puzzle difficulty by collecting data on solving behaviour as well as reported difficulty.

\section{Methodology}

\subsection{SAT Problem Formulation and Metrics}
\label{sec:formulation}

Formally, a Nonogram is defined on an $R \times C$ grid. Each row $r$ is associated with an ordered sequence of positive integers
\[
\mathbf{a}^{(r)} = (a_1, \dots, a_k),
\]
where $a_i$ denotes the length of the $i$-th contiguous block of filled cells. Blocks must appear in order and be separated by at least one empty cell. Column clues $\mathbf{b}^{(c)}$ are defined analogously. A valid solution is a binary matrix $X \in \{0,1\}^{R \times C}$ satisfying all row and column constraints.

We encode this constraint satisfaction problem as a Boolean satisfiability (SAT) instance in conjunctive normal form (CNF). The encoding introduces two types of Boolean variables. For each cell $(r,c)$ we create a variable $x_{r,c}$ indicating whether the cell is filled. For each block $i$ in a row or column and each feasible starting position $p$, we introduce a block-start variable $s_{i,p}$, where $s_{i,p}=1$ iff block $i$ begins at position $p$. Each block must select exactly one start position.

Feasible start positions are computed using tight bounds derived from a packed configuration of all blocks. Let a line have length $N$ and block lengths $(L_1,\dots,L_k)$. The earliest start of block $i$ is
\[
\min_i = \sum_{j<i}(L_j + 1).
\]
Let
\[
T = \sum_{j=1}^{k} L_j + (k-1)
\]
be the minimum length required for all blocks. The remaining slack is $S = N - T$, yielding the latest start $\max_i = \min_i + S$. Restricting $p$ to $[\min_i,\max_i]$ eliminates structurally impossible placements before CNF construction.

The constraints enforce exactly-one placement of each block, ordering between consecutive blocks, coverage of cells by selected blocks, and justification of filled cells. Row and column constraints share the same cell variables, ensuring global consistency. A full description of the clause construction is given in Appendix~\ref{appendix:encoding}.

Let $N$ denote line length and $k$ the number of blocks. Each line introduces $O(kN)$ auxiliary variables and $O(kN^2)$ clauses in the worst case due to pairwise and ordering constraints. For the puzzle sizes considered in our experiments (up to $10\times10$ grids), the resulting CNF instances remain compact and are solved within milliseconds by modern CDCL solvers.

We adopt a block-start encoding rather than a purely cell-based formulation because it exposes explicit block-placement decisions and yields stronger propagation during search.

\subsubsection{Solver-Derived Difficulty Metrics}
\label{sec:metrics}

We collect the following metrics from the SAT solver for each puzzle: the number of decisions, which occur when the solver chooses a variable value arbitrarily (based on a heuristic) because it cannot yet deduce the value logically; the number of propagations, which occur when the solver deduces a variable assignment that must be true to satisfy a clause; and the number of conflicts, which occur when a clause becomes false under the current partial assignment. We treat these statistics as measures of solver search effort and investigate whether higher values correspond to greater human-perceived difficulty.

All solver-metric results reported in this paper use MiniSat 2.2 \cite{een_extensible_2004}, as its minimal preprocessing preserves a clear correspondence between CDCL search behaviour and interpretable metrics such as conflicts, decisions, and propagations. Glucose 4.2 \cite{audemard_glucose_2018} is used only as a secondary solver, to verify that these trends are not specific to a single CDCL implementation. We do not use heavily preprocessing solvers such as CaDiCaL \cite{biere_cadical_2024} for behavioural comparisons, as aggressive simplification may significantly alter solver statistics and obscure their relationship with human solving processes.

\subsection{Experimental Design}

The experiment was designed to collect both fine-grained behavioural interaction data and subjective difficulty judgments while controlling for prior experience and learning effects.

\subsubsection{Participants and Recruitment}

Participants were recruited between January 2026 and May 2026 through email outreach and flyers posted in university buildings. Eligible participants were students registered at a North American research university.

The study protocol was reviewed and approved by the ethics board of a North American research university. Most participants completed the study in person, although a smaller number participated remotely using the same web-based experimental platform and study procedure. Gender and other demographic characteristics beyond age eligibility and academic program were not collected.

\subsubsection{Procedure}

Participants were asked to complete the study individually using a web-based platform. Each session consisted of a tutorial, a warmup phase, a pre-survey, three experimental puzzles, and a post-survey.

The session began with a short tutorial video (approximately two minutes) introducing the rules of Nonograms and demonstrating the user interface. Participants then completed a warmup puzzle (maximum five minutes) to familiarize themselves with the interface and available features. Data from this phase were not included in the analysis.

Next, participants completed a pre-survey collecting background information about their experience with Nonograms and other logic puzzles, including prior exposure to puzzle sizes, frequency of puzzle play, and self-reported skill levels on a 1--10 scale.

Participants then attempted three experimental puzzles sequentially. During each puzzle they could toggle cells, cross cells, undo actions, reset the grid, request hints, submit solutions for verification, or skip the puzzle if they were unable to complete it. Immediately after each puzzle (whether solved or skipped), participants reported the perceived difficulty on a 1--5 scale (with an N/A option for skipped puzzles) and indicated how frequently they relied on guessing rather than logical deduction.

After completing all puzzles, participants filled out a post-survey. In this survey, they were allowed to revise their difficulty ratings and provided open-ended explanations for their judgments, additional self-reports of guessing frequency, descriptions of strategies used, and optional comments about perceived puzzle difficulty.

\subsubsection{Puzzle Selection}
\begin{figure}[ht]
    \centering
    \includegraphics[width=1\linewidth]{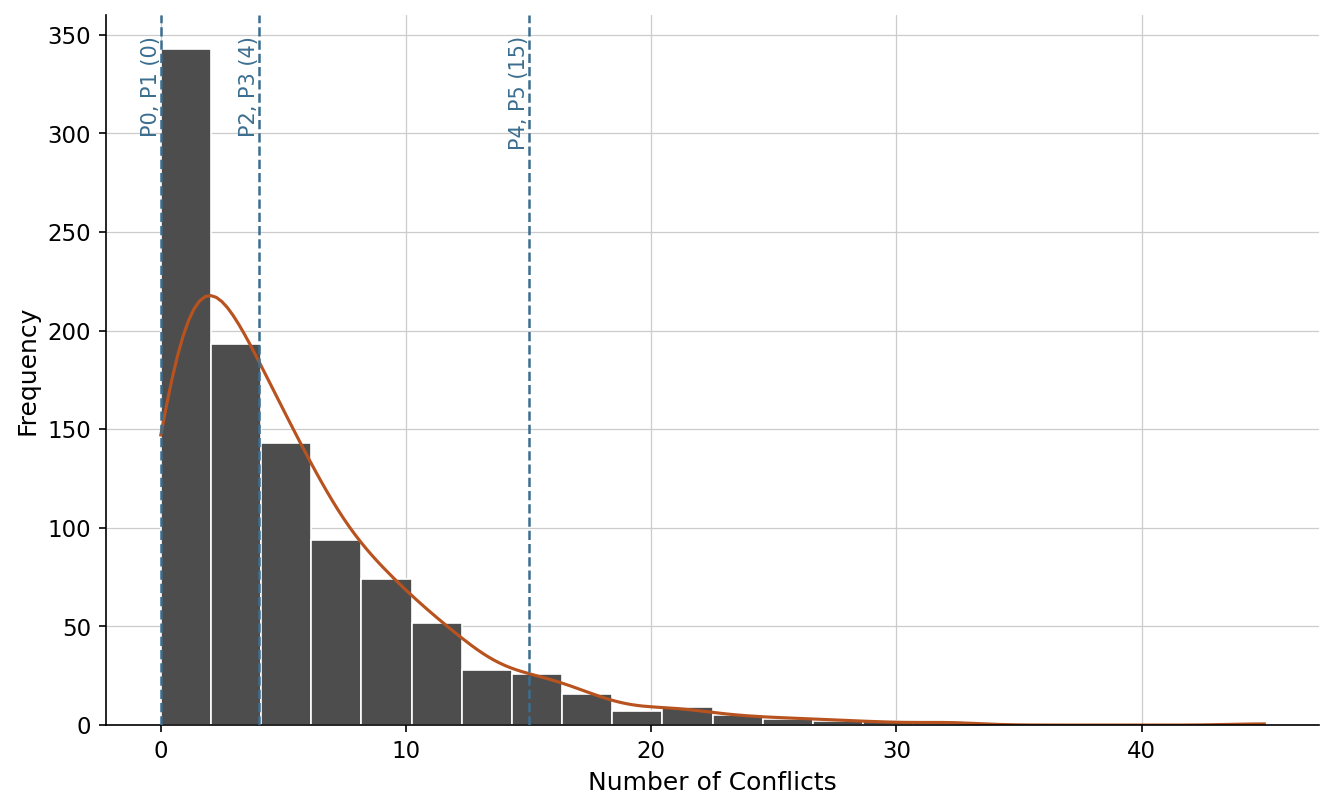}
    \caption{Distribution of number of conflicts over 1,000 randomly generated qualifying puzzles.}
    \label{fig:solver-conflict-dist}
\end{figure}

We generated 1,000 random uniquely solvable $10 \times 10$ Nonograms with 50\% filled cells. Prior computational work suggests that Nonogram solving behaviour varies with both grid size and filled-cell density: substantial changes in the number of unresolved cells have been observed when either parameter is varied \cite{batenburg_reasoning_2008}, and density has been identified as a major factor in inference difficulty \cite{foote_nonogram_2025}. We therefore fixed both grid size and density to reduce these structural sources of variation and make the candidate puzzles more comparable. We then ranked the puzzles by MiniSat 2.2 conflict count and selected six: two with the lowest number of conflicts, two with conflict counts near the median, and two with the highest number of conflicts, excluding the top 5\% (see Figure~\ref{fig:solver-conflict-dist}). This produced a small set spanning low, medium, and high solver-measured difficulty without relying on extreme outliers.

\subsubsection{Measured Variables}

The experiment recorded both behavioural interaction data and subjective self-reports. Raw interactions recorded include total solving time, cell updates, undo and reset operations, hint requests, incorrect solution submissions, and whether the puzzle was solved or skipped. All interaction events were timestamped, allowing reconstruction of complete solving trajectories.

Subjective measures consisted of perceived difficulty ratings (1--5 scale) and self-reported frequency of guessing. The post-survey additionally collected qualitative explanations of difficulty judgments and descriptions of solving strategies, enabling exploratory analysis of reasoning styles.

Finally, the pre-survey collected measures of prior Nonogram experience, general logic puzzle experience, and self-rated skill levels. These variables allow us to control for individual differences when comparing human performance with solver-derived difficulty metrics.

\subsubsection{Implementation Details}

The web-based experimental platform was implemented using Next.js and React, with participant logs stored as newline-delimited JSON files. Nonograms were encoded in Python and SAT solving was performed through PySAT 1.8.dev23 \cite{ignatiev_pysat_2018}. As outlined in Section~\ref{sec:metrics}, all metrics reported in this paper come from MiniSat 2.2, with Glucose 4.2 used only for the secondary verification; CaDiCaL 1.9.5 was run solely to independently confirm puzzle solvability and did not contribute to any reported metric.

\section{Results}

A total of 68 participants completed the study; 1 requested withdrawal, leaving 67 participants and 201 puzzle attempts in the final analysis. Each participant attempted three of the six puzzles. We enumerated all 120 possible ordered selections of three distinct puzzles and assigned participants unique sequences without replacement; therefore, 67 of the 120 sequences were used. This varied both puzzle subset and presentation order, helping reduce systematic order effects. We use participants' revised, final difficulty ratings from the post-survey. Because each participant rates only 3 of the 6 puzzles, we fit a Bradley-Terry model to obtain comparable adjusted rankings across all six puzzles.

\begin{table}[ht]
\centering
\small
\setlength{\tabcolsep}{4pt}
\begin{tabular}{lrr}
\toprule
\textbf{SAT metric} & \textbf{$\rho$}  & \textbf{$p$} \\
\midrule
Decisions & -0.257 & 0.623 \\
Propagations & -0.257 & 0.623 \\
Conflicts & +0.000 & 1.000 \\
\bottomrule
\end{tabular}
\caption{Spearman correlation between Bradley-Terry-adjusted puzzle difficulty ranking and SAT solver metrics.}
\label{tab:bt-vs-sat}
\end{table}

\textbf{RQ1:} We conducted a Spearman rank correlation test between the BT-adjusted participant-reported difficulty rankings and SAT solver metric rankings, and found no significant correlation at the puzzle level (Table~\ref{tab:bt-vs-sat}).

\subsection{Expertise}

Participants reported expertise on six items: Nonogram skill (1--10), general puzzle skill (1--10), how often they play Nonograms, how often they play puzzles, the range of Nonogram sizes they've tried, and the range of logic-puzzle types they know.

PCA on the six dimensions of participant self-reported expertise shows that PC1 alone explains 56\% of the variance, and all six items load positively, suggesting that there is a single dominant dimension. Factor analysis scores agree closely with the naive z-mean composite, obtained from averaging the z-scores of each of the six dimensions per participant. For simplicity of interpretation, we choose to represent expertise with z-mean in the analysis that follows.

\begin{figure}[ht]
    \centering
    \includegraphics[width=1\linewidth]{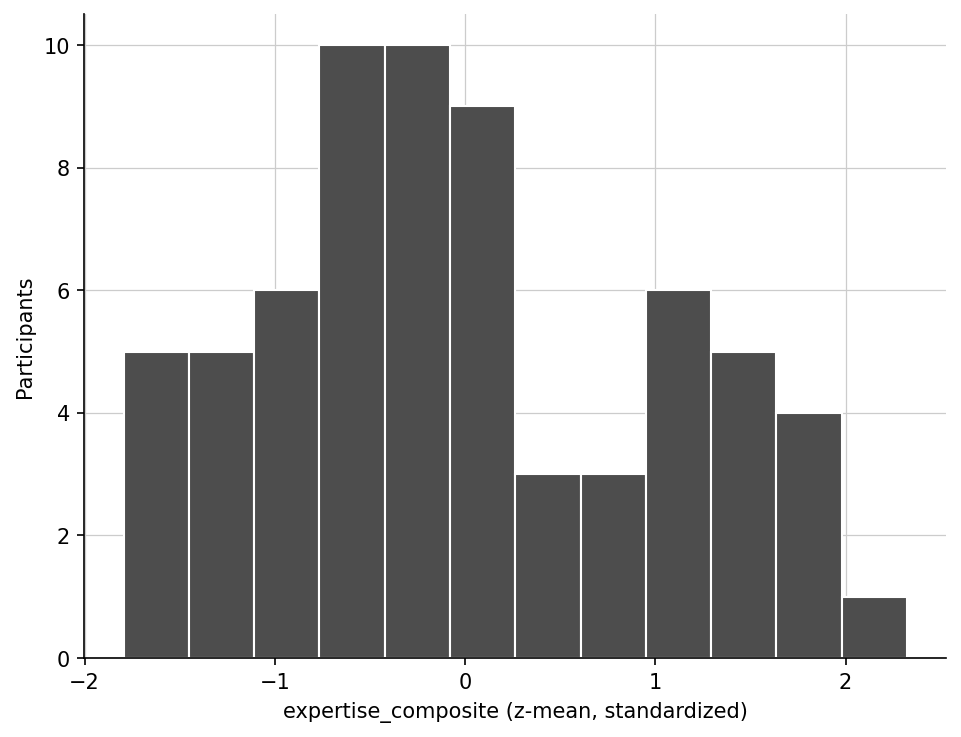}
    \caption{The distribution of the z-mean expertise score across participants.}
    \label{fig:expertise-composite-dist}
\end{figure}

Conducting a Spearman test between participant expertise composite and their mean final difficulty, we find no statistically significant relationship ($\rho = -0.094$, $p = 0.451$). However, fitting a crossed-random-effects linear mixed model (LMM), with random intercepts for both participant ID and puzzle ID simultaneously, we discover that expertise significantly moderates the relationship between SAT solver metrics and reported difficulty through a negative coefficient (see Table~\ref{tab:moderation-interaction}). That is, more expert participants' difficulty ratings become progressively less sensitive to how hard the puzzle is as measured by SAT-solver metrics.

\begin{table}[ht]
\centering
\small
\setlength{\tabcolsep}{4pt}
\begin{tabular}{lccc}
\toprule
\textbf{SAT metric} & \textbf{Metric $\beta$} & \textbf{Expertise $\beta$} & \textbf{Interaction $\beta$} \\
\midrule
Decisions & -0.0007 & +0.0875 & -0.0096* \\
Propagations & +0.0001 & +0.2053 & -0.0004** \\
Conflicts & +0.0067 & +0.0891 & -0.0247** \\
\bottomrule
\end{tabular}
\caption{Moderation of the SAT-metric $\to$ difficulty relationship by participant expertise: fixed effects from a crossed-random-effects LMM (difficulty $\sim$ metric $\times$ expertise, with crossed random intercepts for participant and puzzle). * $p<.05$, ** $p<.01$.}
\label{tab:moderation-interaction}
\end{table}

\subsection{Behavioural Data}

We extract the following signals from data collected as the participants solve the puzzles:

\begin{itemize}
  \item \textbf{Time to solve:} the elapsed time it takes for a participant to solve the puzzle (submit a solution that is correct). If a participant fails to solve the puzzle (skipped), the time to solve is the total amount of time that the participant spent before skipping.
  \item \textbf{Pause count:} the number of times a participant paused in the process of solving a puzzle. A pause is any gap in interaction that exceeds $2.36$s, the equal-posterior threshold between components in the gap length distribution (see Appendix~\ref{appendix:pauses}).
  \item \textbf{Error count:} the number of times a participant filled a cell black where the correct solution is white.
  \item \textbf{Hint count:} the number of hints the participant used.
\end{itemize}

\begin{table}[ht]
\centering
\small
\setlength{\tabcolsep}{4pt}
\begin{tabular}{lrr}
\toprule
\textbf{Outcome} & \textbf{$\rho$} & \textbf{$p$} \\
\midrule
Mean time to solve (s) & -0.541 & $<$.001 \\
Mean pause count & -0.507 & $<$.001 \\
Mean error count & -0.407 & $<$.001 \\
Mean hints used & -0.583 & $<$.001 \\
\bottomrule
\end{tabular}
\caption{Spearman correlation between participant expertise (composite score) and behavioural outcomes, aggregated to participant-level means.}
\label{tab:expertise-vs-outcomes}
\end{table}

A second, participant-level Spearman correlation test shows a strong relationship between expertise and behaviour (see Table~\ref{tab:expertise-vs-outcomes}): higher expertise scores were associated with faster solving, fewer hints and pauses, and fewer errors; this further validates our approach for constructing the expertise composite.

\begin{figure}[ht]
    \centering
    \includegraphics[width=1\linewidth]{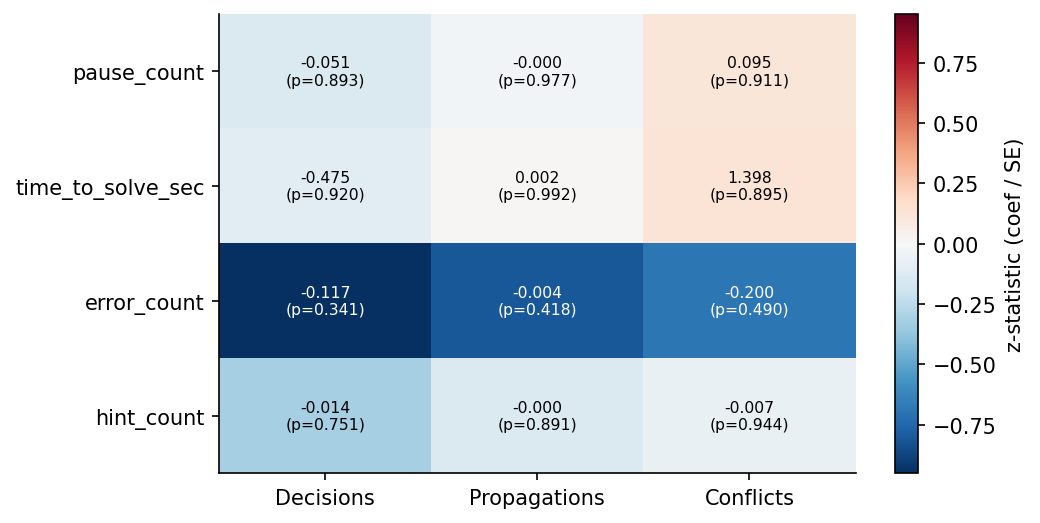}
    \caption{Behavioural signals regressed on SAT solver metrics: fixed effects from a crossed-random-effects LMM (behavioural signal $\sim$ metric, with crossed random intercepts for participant and puzzle).}
    \label{fig:behaviour-sat-relation}
\end{figure}

\textbf{RQ2:} Fitting a crossed-random-effects LMM for each behavioural signal and SAT metric pair, we find no significant correlation between any particular recorded participant behaviour and solver metric (see Figure~\ref{fig:behaviour-sat-relation}).

\subsection{Survey Responses}

For each puzzle, participants were asked to respond to two questions: \textit{``Why did you rate this difficulty? What made it easy/hard?''} and \textit{``What strategies did you use when solving the puzzles?''} They also provided any additional general comments. Using the Claude Sonnet 4.6 language model (an approach partially supported by \cite{xiao_supporting_2023}), the responses were coded using two categories of labels: thirteen difficulty themes and eight strategy themes. The full codebook, along with an example occurrence for each code, is included in Appendix~\ref{appendix:codebook}.

\begin{figure}[ht]
    \centering
    \includegraphics[width=1\linewidth]{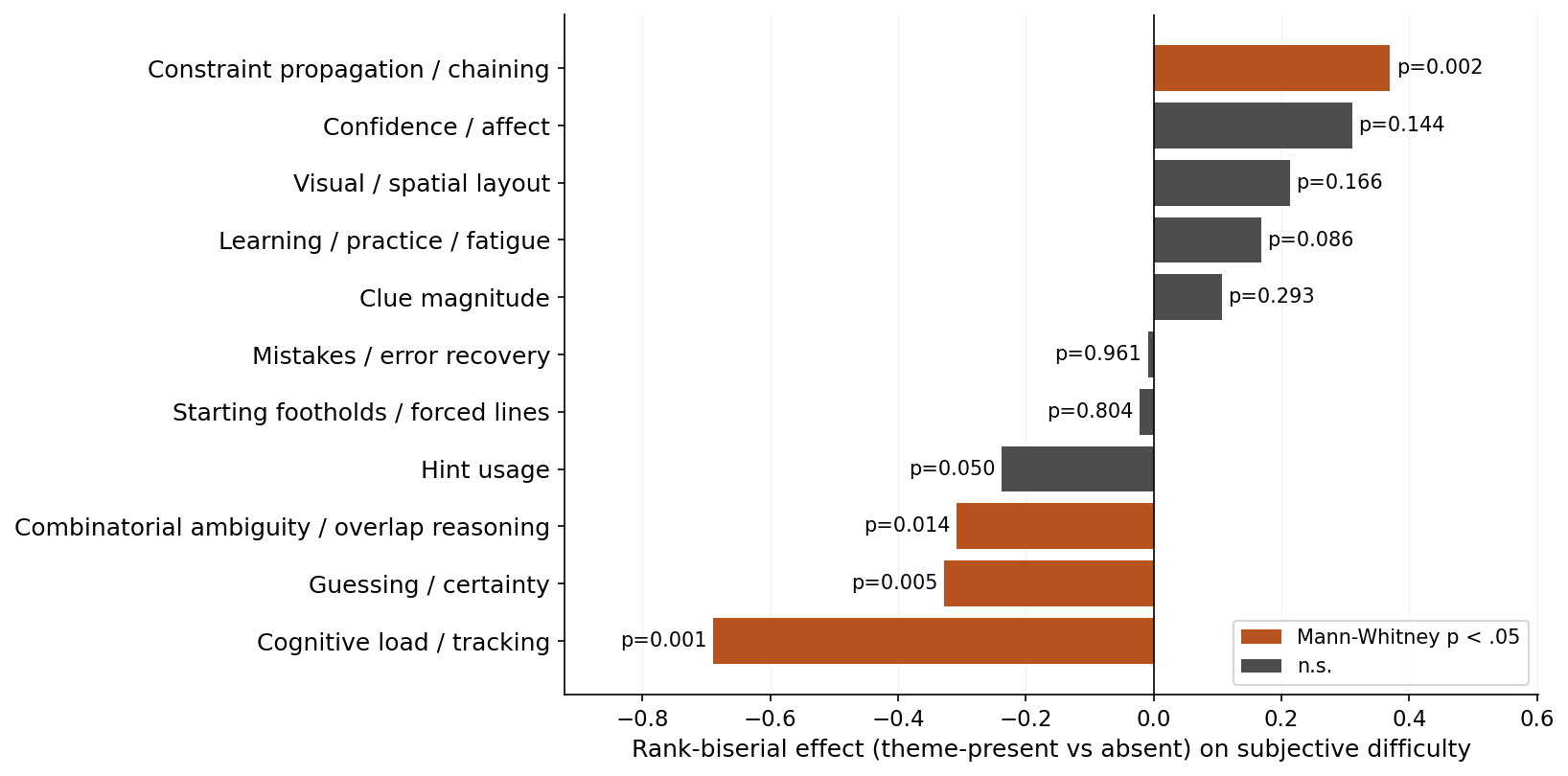}
    \caption{Relationship between each difficulty theme and participant-reported difficulty.}
    \label{fig:theme-difficulty}
\end{figure}

We conducted rank-biserial correlation on the various themes against participant-rated difficulty; Figure~\ref{fig:theme-difficulty} shows that constraint propagation is negatively correlated with subjective difficulty, whereas ambiguity, guessing, and cognitive load are positively correlated with subjective difficulty.

\section{Discussion}

We acknowledge that the investigations into puzzle-level correlations are exploratory due to the small sample size of 6 puzzles, and the puzzle assignments are somewhat unbalanced (28--43). Notwithstanding these limitations, the lack of evidence for SAT solver metrics predicting human-perceived difficulty (as reported or proxied by behavioural signals) suggests that human and solver difficulty arise from different underlying mechanisms.

\begin{figure}[ht]
    \centering
    \includegraphics[width=1\linewidth]{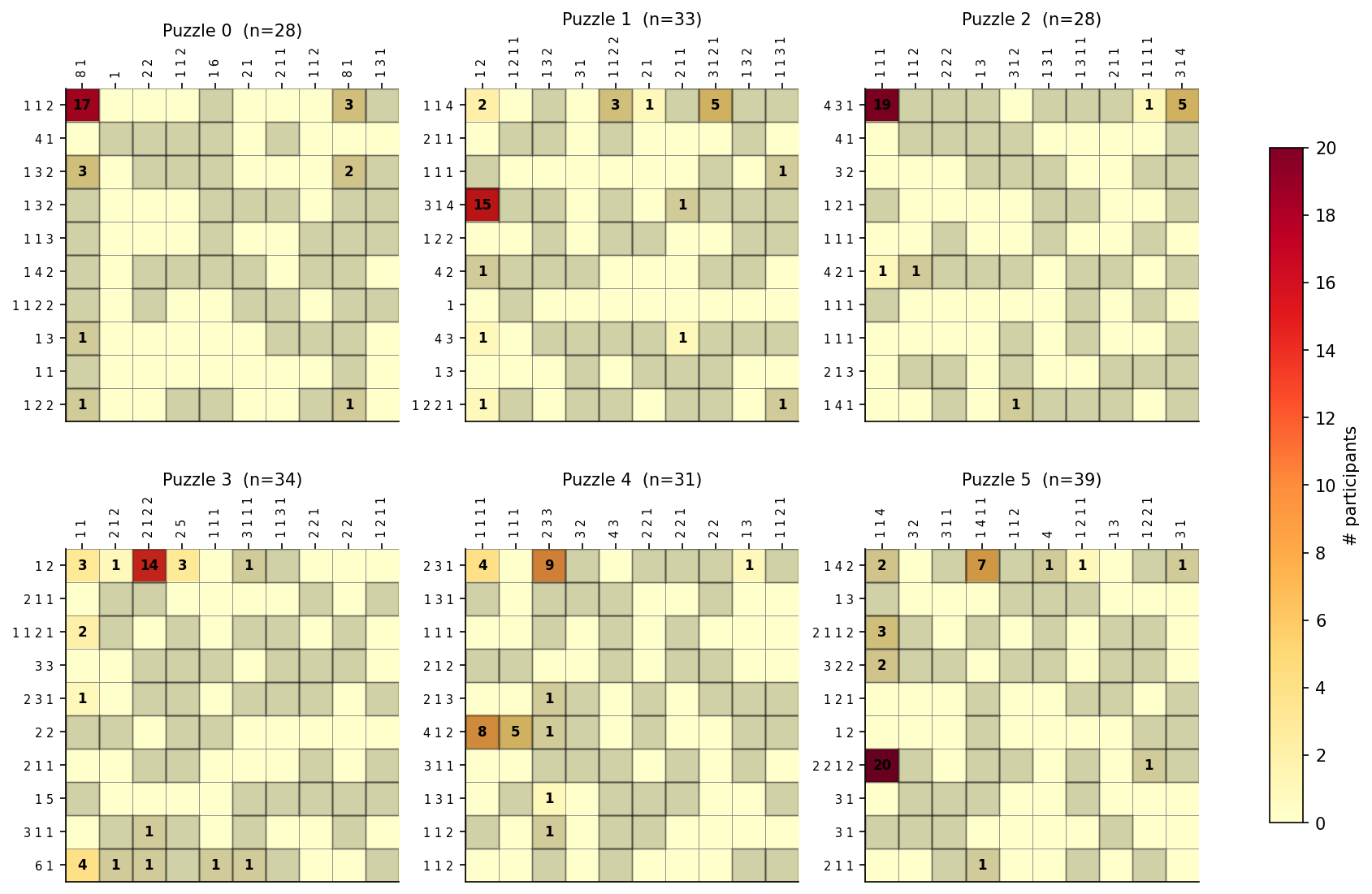}
    \caption{Heatmap of the first cell acted upon for each puzzle, by frequency.}
    \label{fig:solve-heatmap}
\end{figure}

Prior Sudoku research has proposed that clarity and intuitiveness are critical components of perceived difficulty \cite{wang_rating_2012}. Relatedly, humans have been shown to prefer constraint propagation and to be averse to backtracking \cite{pelanek_difficulty_2011}. Our results support this claim: participants who mentioned constraint propagation rated their puzzles as meaningfully easier, while guessing, ambiguity, and cognitive load--associated with backtracking--were more common among puzzles rated as harder. Further, \cite{simonis_sudoku_2005} suggest that humans apply complex propagation schemes; these align with strategies that puzzle-hosting websites commonly share as tips for players to improve, several of which were reported by our own participants. The strategy theme mentioned most frequently at 46\%, the ``foothold''--that is, a line whose configuration can be fully determined from its clues alone--is visible in a heatmap of the first actions that players take (see Figure~\ref{fig:solve-heatmap}).

We posit that this causes a meaningful divide between solving processes of the human and the SAT solver. A human solver focuses on a ``strategy'' and prioritizes using advanced propagation schemes that combine several constraints at once. In contrast, because we only encoded the basic rules of the Nonogram, reasoning strategies that involve multiple constraints are not available to SAT solvers, which are limited to unit propagation on individual clauses. However, CDCL-backed solvers like MiniSat 2.2 learn joint patterns of clauses after encountering conflicts while solving the problem, enriching their understanding; this may be analogous to the process by which beginner players discover and start to implement strategies during solving. Indeed, Table~\ref{tab:moderation-interaction} shows evidence that expert difficulty ratings diverge further from solver ratings than beginners' ratings do. Thus, it is possible that solvers would be better at capturing the difficulties encountered by human players with more propagation-complete encodings, as higher expressiveness may better capture propagation schemes used by humans.

\section{Conclusion}

This exploratory study found that SAT solver metrics largely do not align with how humans experience and solve Nonograms; trajectory and survey response analysis revealed recurring human solving strategies that solver metrics do not capture. We find some evidence that expertise moderates the relationship between reported difficulty and solver metrics, and hypothesize that this reflects participants' growing understanding of advanced propagation strategies. We call for further investigation into whether the expressiveness of CSP encodings of puzzles affects the efficacy of constraint solver metrics as difficulty-tracking methods. Additionally, future work could extend this study to a larger scale with more puzzle samples, as well as to other logic puzzles (e.g., Sudoku, Norinori). More broadly, our results invite a reassessment of the role of computational proxies in predicting difficulty for puzzles.

\section*{Acknowledgements}

We thank Juliana Zhang for developing the frontend of the experimental platform. We thank Riyad Valiyev for assisting with running study sessions and exploratory analysis of participant text responses. We also thank the study participants for their time and feedback.

\appendix

\section{Detailed SAT Encoding}
\label{appendix:encoding}

This section provides the detailed clause construction for the SAT encoding used in this study.

For each cell $(r,c)$, we introduce a Boolean variable $x_{r,c}$ indicating whether the cell is filled. For each block $i$ and feasible start position $p$, we introduce a variable $s_{i,p}$.

\paragraph{Exactly-one placement.}
Each block must be placed exactly once. Let $\{s_{i,p}\}_p$ denote the set of feasible start variables for block $i$. We enforce
\[
\bigvee_p s_{i,p}
\]
and, for all $p \neq q$,
\[
\neg s_{i,p} \vee \neg s_{i,q}.
\]

\paragraph{Block ordering.}
To enforce separation between consecutive blocks, if block $i$ starts at $p$, block $i+1$ must start at position at least $p + L_i + 1$. For any invalid pair $(p,q)$ with $q < p + L_i + 1$, we add
\[
\neg s_{i,p} \vee \neg s_{i+1,q}.
\]

\paragraph{Coverage constraints.}
If block $i$ starts at $p$, all cells in its span must be filled:
\[
s_{i,p} \rightarrow x_j
\quad \text{for } j \in [p,p+L_i-1],
\]
encoded as
\[
\neg s_{i,p} \vee x_j.
\]

\paragraph{Cell justification.}
Any filled cell must be covered by at least one block placement:
\[
x_j \rightarrow \bigvee_{(i,p)\text{ covering } j} s_{i,p},
\]
which is encoded as
\[
\neg x_j \vee s_{i_1,p_1} \vee \dots \vee s_{i_m,p_m}.
\]
If no block placement can cover a cell $j$, the clause $\neg x_j$ is added.

\paragraph{Empty lines.}
If a line contains no blocks, all cells in that line are constrained to be empty.

\section{Identifying Pauses}
\label{appendix:pauses}

We separate all gaps in interaction into three components using a Gaussian mixture model ($\Delta\mathrm{BIC}=7387.1$ against one component). The first component, with mean $0.01$s, results from the interface allowing participants to drag across cells to fill them consecutively. We separate the rest of the interactions into two components to distinguish smooth logic chaining from blocks in reasoning or deliberation. The pause threshold is therefore selected to be $2.36$s, the equal-posterior threshold between components 2 and 3.

\begin{figure}[ht]
    \centering
    \includegraphics[width=1\linewidth]{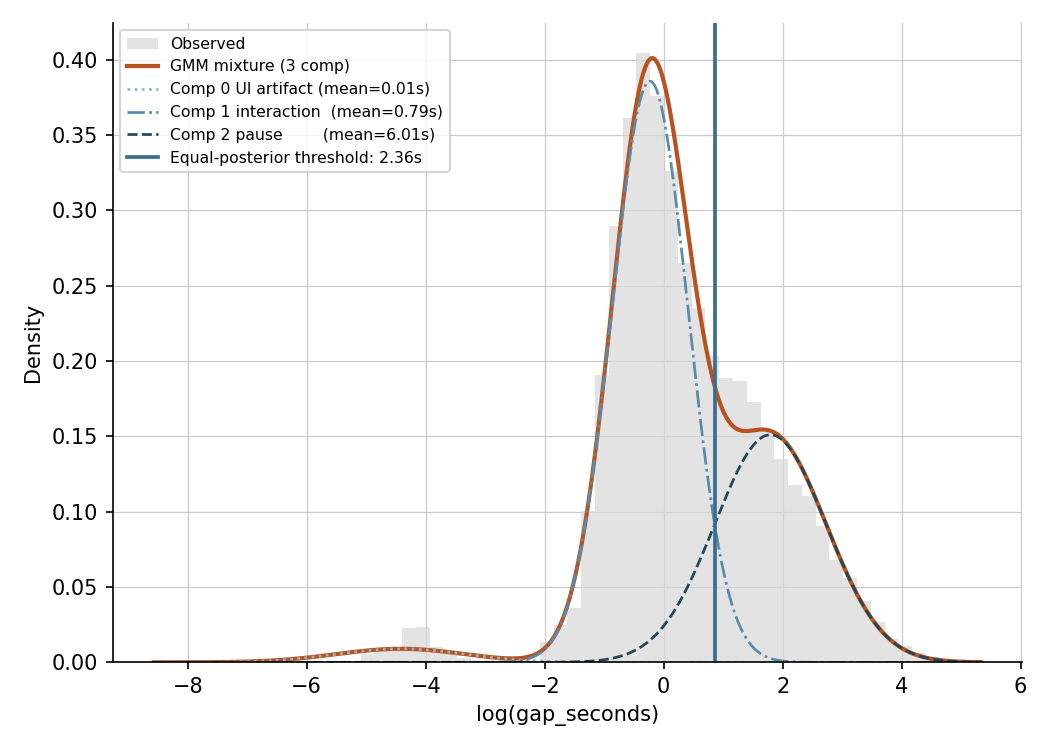}
    \caption{Log-interval distribution of interaction gaps recorded from participant behaviour during puzzle solving, with a three-component GMM.}
    \label{fig:gap-log-intervals}
\end{figure}

\section{Codebook}
\label{appendix:codebook}

Table~\ref{tab:codebook-examples} gives the codes used in the qualitative analysis, with an example occurrence of each code from the dataset.

{\small
\renewcommand{\arraystretch}{1.0}
\setlength{\tabcolsep}{4pt}
\begin{longtable}{@{}llp{2.4cm}p{8.8cm}@{}}
\caption{Difficulty themes and strategy taxonomy, with example responses.}
\label{tab:codebook-examples}\\
\toprule
\textbf{Category} & \textbf{Code} & \textbf{Theme / Strategy} & \textbf{Example} \\
\midrule
\endfirsthead
\toprule
\textbf{Category} & \textbf{Code} & \textbf{Theme / Strategy} & \textbf{Example} \\
\midrule
\endhead
\bottomrule
\endfoot
\textbf{Difficulty} & \texttt{FOOT} & Starting footholds / forced lines & ``Not a lot of freebies'' \\[3pt]
 & \texttt{CLUE} & Clue magnitude & ``Clue number too small'' \\[3pt]
 & \texttt{PROP} & Constraint propagation / chaining & ``easier to solve using elimination'' \\[3pt]
 & \texttt{AMBIG} & Combinatorial ambiguity / overlap reasoning & ``Multiple intermediate steps seemed possible, but only one worked in the end'' \\[3pt]
 & \texttt{GUESS} & Guessing / certainty & ``never need to guess'' \\[3pt]
 & \texttt{HINT} & Hint usage & ``I use many hints.'' \\[3pt]
 & \texttt{ERR} & Mistakes / error recovery & ``I got a little stuck but maybe that's because I misread some of the numbers'' \\[3pt]
 & \texttt{LEARN} & Learning / practice / fatigue & ``easier than first one'' \\[3pt]
 & \texttt{LOAD} & Cognitive load / tracking & ``Keeping track of what has been filled out made things overall harder.'' \\[3pt]
 & \texttt{VIS} & Visual / spatial layout & ``In a lot of other nonograms, they usually form a picture so you kind of get a free hint as to what the general shape should look like but these ones were more random'' \\[3pt]
 & \texttt{TIME} & Time pressure & ``Guessed because of time constraints.'' \\[3pt]
 & \texttt{UI} & Interface friction & ``undo button should not be the same color and right beside the reset button'' \\[3pt]
 & \texttt{AFF} & Confidence / affect & ``I never got stuck and felt confident the entire time.'' \\[3pt]
\midrule
\textbf{Strategy} & \texttt{S\_FORCED} & Forced-line solving & ``determine the fixed lines first'' \\[3pt]
 & \texttt{S\_CONSTR} & Most-constrained-first ordering & ``Locate the largest number block first.'' \\[3pt]
 & \texttt{S\_OVERLAP} & Overlap analysis & ``finding overlaps'' \\[3pt]
 & \texttt{S\_EDGE} & Edge / anchor exploitation & ``fixed arrangements from start/end positions, see possible solutions with remaining white space'' \\[3pt]
 & \texttt{S\_CROSS} & Row/column cross-referencing & ``Trying to compare values across rows and columns to see where the hints matched.'' \\[3pt]
 & \texttt{S\_NEG} & Negative marking (X-ing whites) & ``To fulfill the known lines first, and then along the guessing, I can cross some connected cells.'' \\[3pt]
 & \texttt{S\_TRIAL} & Trial-and-error / contradiction & ``logical deduction, trial and error'' \\[3pt]
 & \texttt{S\_HINT} & Hint-as-tool & ``I feel like it's better to use hints especially at the very beginning, but I had a lot left when finished.'' \\[3pt]
\end{longtable}
}

\newpage
    \bibliographystyle{plain}
    \bibliography{ref}

\end{document}